\documentclass[aps,prx,reprint]{revtex4-2}
\usepackage[utf8]{inputenc}
\usepackage{microtype}
\usepackage{xspace}
\usepackage{mathrsfs}
\usepackage{amsfonts,mathtools,mathrsfs}
\usepackage{bm}
\usepackage{xcolor}
\usepackage[colorlinks=true, linkcolor = blue, hyperfootnotes=false, citecolor = blue]{hyperref}

\newcommand{\str}{\varepsilon}

\newcommand{\abs}[1]{\left\lvert #1 \right\rvert}

\newcommand{\scalar}[1]{\left\langle{#1}\right\rangle}
\newcommand{\dd}{\mathrm{d}}
\renewcommand{\t}{\mathbf}
\newcommand{\gt}{\bm}
\renewcommand\[{\begin{equation}}
\renewcommand\]{\end{equation}}

\newcommand{\comment}[1]{#1}
\newcommand{\elltwo}{\mathbb{E}}
\newcommand{\ellone}{\mathbb{K}}

\begin{document}
\title{Curved-crease origami for morphing metamaterials}
\author{Asma Karami}
\author{Adam Reddy}
\author{Hussein Nassar}\email{Corresponding author: nassarh@missouri.edu}
\affiliation{Department of Mechanical and Aerospace Engineering, University of Missouri, Columbia, MO 65211, USA}

\begin{abstract}
We find a closed-form expression for the Poisson's coefficient of curved-crease variants of the ``Miura ori'' origami tessellation. This is done by explicitly constructing a continuous one-parameter family of isometric piecewise-smooth surfaces that describes the action of folding out of a reference state. The response of the tessellations in bending is investigated as well: \comment{using a numerical convergence scheme,} the effective normal curvatures under infinitesimal bending are found to occur in a ratio equal and opposite to the Poisson's coefficient. These results are the first of their kind and, by their simplicity, should provide a fruitful benchmark for the design and modeling of curved-crease origami and compliant shell mechanisms. Here, the developed methods are used to design a curved crease 3D morphing solid with a tunable self-locked state.

\end{abstract}
\maketitle

The properties of architected materials, and metamaterials, are just as much, if not more, a product of geometry and spatial layout as they are a product of the properties of the raw materials they are made of. Here is a striking example from the classical theory of composites~\cite{Milton2002}: the effective Young's modulus of an isotropic porous plate does not depend on the Poisson's coefficient of the raw material and only depends on the spatial distribution of the pores~\cite{Day1992}. Moreover, for highly porous plates, the effective Poisson's coefficient is a pure product of geometry and is completely independent of the raw material~\cite{Thorpe1992}.

Origami tessellations, and other architected materials composed of structural elements including bars and plates connected at joints, provide more recent examples where the Poisson's coefficient is a purely geometric construct~\cite{Wang2020, Ren2021}. Such structures are near-mechanisms and possess a soft deformation mode whose effective Cauchy-Green tensor $\t I$ defines an instantaneous (Lagrangian) Poisson's coefficient
\[
    \nu \equiv -\frac{\dd I_{22}/I_{22}}{\dd I_{11}/I_{11}}.
\]
Typically, the soft deformation mode engages rotations at the joints and moves the structural elements as rigid bodies. This makes the computation of $\nu$ particularly straightforward, however algebraically complex~\cite{Schenk2013}. That said, there are highly-compliant structures that cannot be treated in the same fashion, namely as linkages, for their soft deformation modes necessarily involve the bending of the structural elements~\cite{Seffen2012}. Examples include inextensible elastic curves and surfaces. For curved-crease origami~\cite{Demaine2015b}, the difficulty is compound: on one hand, the presence of curved folds couples folding angles to bending in the facets; on the other hand, the inextensible bending within facets still have to be compatible at the curved crease lines.

\begin{figure*}
    \centering
    \includegraphics[width=\linewidth]{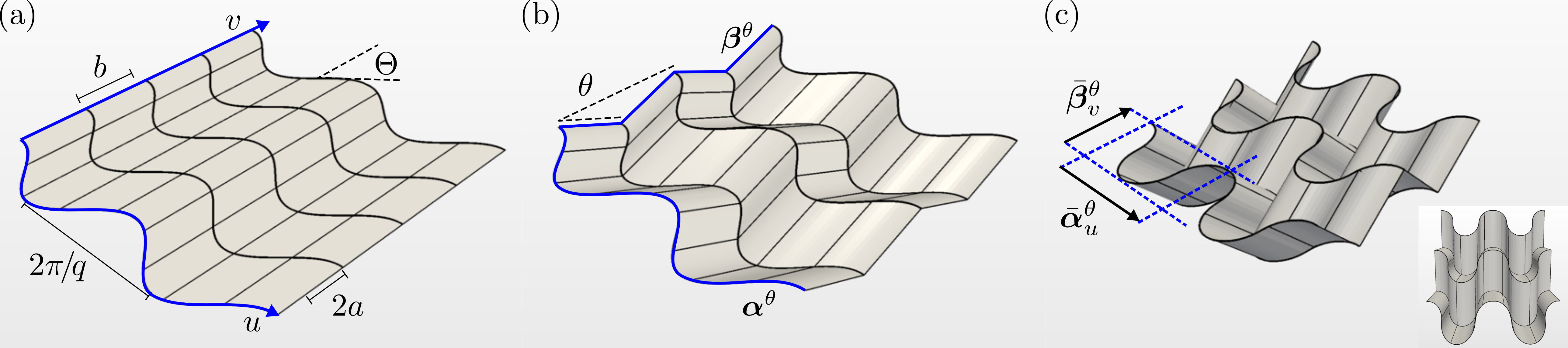}
    \caption{A curved-crease variant of the Miura ori and used notations. (a) Crease pattern in a flat reference configuration: some ruling segments are revealed to improve visibility; thicker curved lines are crease lines. (b) Partially folded state. (c) Fully folded state: $\theta=\Theta$; inset is an axial view that better shows the (locally) flat-folded ruling segments.
    Used values: $a=q=1$; $\Theta=\pi/4$; $b=3$.}
    \label{fig:1}
\end{figure*}

Rigorous results in the field of curved-crease origami are relatively recent even though a characterization of developable surfaces has been known since Euler. Recall that a developable surface is a smooth surface that is locally isometric to a plane, i.e., it is a surface assembled from planar pieces that are bent without stretching or creasing. Euler showed that developable surfaces are ruled: they are composed of straight segments of finite lengths. Furthermore, the ruling is torsal: the tangent plane maintains tangency along each rule segment ~\cite{Fuchs1999}. What was lacking, and is currently actively researched, is an understanding of how the rule segments of different developable facets can fit together along crease lines and still produce a surface that is locally isometric to a plane. Demaine championed efforts in this area building on previous work by Huffman~\cite{Huffman1976a, RichardDuksKoschitz2014}. With collaborators, they were able to design, and prove the (mathematical) existence of, multiple curved-crease origami sculptures based on an understanding of what distributions of rule segments are compatible with what curved crease lines~\cite{Demaine2011, Demaine2015a, Demaine2018}. Discretizations, guided by similar considerations, led to fruitful form-finding numerical tools~\cite{Kilian2008a, Tang2016a, Jiang2019, Demaine2023}. Some semi-analytical asymptotic tools are also available for patterns composed of thin facets (or ribbons)~\cite{Dias2012, Dias2012a}; see also~\cite{Mowitz2022}.

In the present paper, a different approach is proposed. Rather than working facet by facet, a global isometry (by immersion) between the planar and folded state is directly constructed. The isometry is in fact one instance of a continuous one-parameter family of isometries for which closed-form expressions are provided. In particular, this allows to compute the Poisson's coefficient $\nu$ of a curved-crease origami tessellation via simple integration. Evidently, such methods are not universal but are applicable for a class of tessellations that, however restricted, is ubiquitous in engineering applications, including curved-crease foldcores used in sandwich panels~\cite{Gattas2015, Du2019, Deng2022}. To present the approach, a curved-crease variant of the Miura ori is adopted as an archetypical example first. Generalizations to other patterns and applications to the design of curved-crease 3D compliant mechanisms follow.

Thus, let $\gt \alpha: u\mapsto (u,a\sin(q u),0)$ describe a sinusoidal curved crease of parameters $(a>0,q>0)$ and let $\gt \beta: v\mapsto (0,v,0)$ describe a straight line. Consider the parametrization of the plane $\t x: (u,v)\mapsto \gt\alpha(u) + \gt\beta(v)$ illustrated in Fig.~\ref{fig:1}a. The purpose is to determine a one-parameter family of parametrized surfaces $\theta\mapsto \t x^\theta$ that are isometric to $\t x$. These surfaces must be piecewise smooth with jumps in the tangent plane located at a series of crease lines $\{v = m b\}$, with $m$ integer and $b$ being a uniform spacing between two consecutive crease lines. The idea is to construct $\t x^\theta$ as a ``surface of translation'', i.e., in the form $\t x^\theta: (u,v)\mapsto \gt\alpha^\theta(u) + \gt\beta^\theta(v)$ where the curves $\gt\alpha^\theta$ and $\gt\beta^\theta$ are deduced from the originals $\gt\alpha$ and $\gt\beta$ by maintaining the isometric character of the deformation. That is, such that
\[\label{eq:isometry}
\begin{split}
    \scalar{\gt\alpha_u^\theta(u),\gt\alpha_u^\theta(u)} &= \scalar{\gt\alpha_u(u),\gt\alpha_u(u)} = 1 + a^2q^2\cos^2(qu),\\
    \scalar{\gt\beta_v^\theta(v),\gt\beta_v^\theta(v)} &= \scalar{\gt\beta_v(v),\gt\beta_v(v)} = 1,\\
    \scalar{\gt\alpha_u^\theta(u),\gt\beta_v^\theta(v)} &= \scalar{\gt\alpha_u(u),\gt\beta_v(v)} = aq\cos(qu),
\end{split}
\]
where subscripts stand for derivatives. Hereafter, it will prove convenient to build $\gt\alpha^\theta$ and $\gt\beta^\theta$ from their respective tangent vectors $\gt\alpha^\theta_u$ and $\gt\beta^\theta_v$ by integration according to
\[
    \gt\alpha^\theta(u)= \int_0^u \gt\alpha^\theta_u(\xi)\dd\xi,\quad
    \gt\beta^\theta(v) = \int_0^v \gt\beta^\theta_v(\zeta)\dd\zeta.
\]
Now intuition and experiment suggest $\gt\beta^\theta$ be a ``zigzag'' of a tangent vector $\gt\beta^\theta_v$ defined piecewise by\comment{
\[
    \gt\beta^\theta_v(v) =
    \begin{cases}
         (0,\cos\theta, \sin\theta) \quad &\text{for} \quad 0\leq v < b,\\
        (0,\cos\theta, -\sin\theta) \quad &\text{for} \quad b\leq v < 2b,
    \end{cases}
\]
and so on, by alternating signs. Then, let
\[
    \gt\alpha^\theta_u(u) = \left(
    \sqrt{1-a^2q^2\cos^2(qu)\tan^2\theta} ,
    aq\frac{\cos(qu)}{\cos\theta}, 
    0
    \right)
\]
so as to fulfil the requirement of $\t x^\theta$ being isometric to~$\t x$, namely equation~\eqref{eq:isometry}}. Furthermore, $\t x^{t=0}=\t x$. Accordingly, $\theta\mapsto\t x^\theta$ parametrizes a continuous isometric deformation out of the planar state~$\t x$; the family is illustrated on Fig.~\ref{fig:1}. The deformation parameter $\theta$ is the acute angle between the zigzag segments and the plane $\{z=0\}$. That angle cannot be arbitrarily large and is bound by a value that corresponds to a maximally-folded state. Indeed, $\t x^\theta$ is well-defined if, for all $u$,
\[
    1-a^2q^2\cos^2(qu)\tan^2 \theta \geq 0,
\]
i.e., if
\[
    \abs{\theta} \leq \frac{\pi}{2}-\arctan(aq) \equiv \Theta.
\]
In other words, angle $\theta$ must remain smaller than the smallest angle that the curved crease $\gt\alpha$ makes with the ruling $\{x=cst\}$, namely $\Theta$. As $\theta$ reaches its upper bound $\Theta$, the triad composed of the tangents to the curved crease and to the two sides of the zigzag folds flat at $u=\pi m/q$, for $m$ integer (Fig.~\ref{fig:1}c).

\begin{figure}
    \centering
    \includegraphics[width=\linewidth]{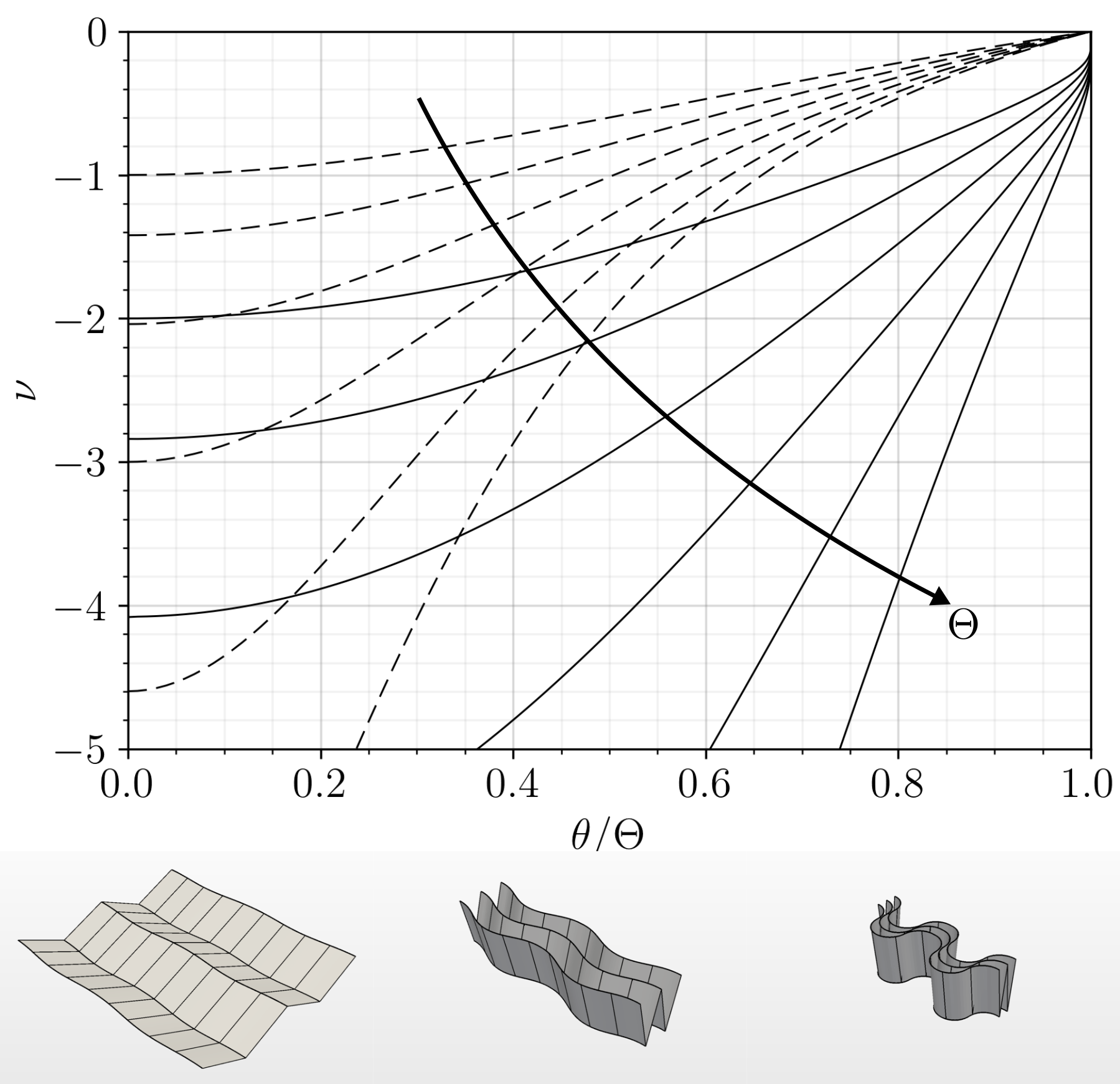}
    \caption{The Poisson's coefficient. Top: Variations of the Poisson's coefficient v.s. the relative folding angle both for the original \cite{Schenk2013} Miura ori (dashed) and curved-crease variant (solid) with equal unit cell dimensions and maximum fold angle $\Theta$; angle $\Theta$ is sampled every $5^\circ$ from $45^\circ$ to $70^\circ$ increasing with the arrow. Bottom: folded states for an extreme value of $\Theta\sim\pi/2$; note how the motion proceeds in one direction then in the other.}
    \label{fig:2}
\end{figure}

It is noteworthy that the folded surfaces $\t x^\theta$ are periodic with a unit cell $[0,2\pi/q]\times[0,2b]$, the same as the original crease pattern. Even though all of these surfaces are mutually isometric, it is possible to define an effective stretch based on the span of the image of the unit cell. Hence, let $\t I^\theta$ be the effective metric of $\t x^\theta$ given by
\[
I^\theta_{11} = \scalar{\bar{\gt\alpha}_u^\theta,\bar{\gt\alpha}_u^\theta},\quad
I^\theta_{22} = \scalar{\bar{\gt\beta}_v^\theta,\bar{\gt\beta}_v^\theta},\quad
I^\theta_{12} = \scalar{\bar{\gt\alpha}_u^\theta,\bar{\gt\beta}_v^\theta},
\]
where $\bar{\gt\alpha}_u^\theta\equiv(q/2\pi)\int_0^{2\pi/q}{\gt\alpha}_u^\theta\dd u$ and $\bar{\gt\beta}_v^\theta\equiv(1/2b)\int_0^{2b}{\gt\beta}_v^\theta\dd v$. It is easy to check that
\[
I^\theta_{11} = \frac{4}{\pi^2}\elltwo^2(\mu),\quad I^\theta_{22} = \cos^2\theta,\quad I^\theta_{12} = 0,
\]
where $\elltwo(\mu)$ is the complete elliptic integral of the second kind evaluated at $\mu=a^2q^2\tan^2\theta=\tan^2\theta/\tan^2\Theta$. In particular, at $\theta=0$, $\t I^{\theta=0}=\t I$ is the identity. In conclusion of this section, \comment{letting $\ellone(\mu)$ be the complete elliptic integral of the first kind,} the Poisson's coefficient is
\[
    \nu(\theta) \equiv -\frac{\dot I_{22}^\theta/I_{22}^\theta}{\dot I_{11}^\theta/I_{11}^\theta} 
    = \frac{\tan^2\theta}{1+\tan^2\theta}\frac{\elltwo(\mu)}{\elltwo(\mu)-\ellone(\mu)}
\]
and is negative. The variations of the Poisson's coefficient of curved-crease variants of the Miura-ori are illustrated on Fig.~\ref{fig:2}~(top). It is seen that the curved creases increase the Poisson's coefficient, in absolute value, by a factor of 2 for comparable maximum angles $\Theta$ and folding fractions~$\theta/\Theta$. Note that for larger $\Theta$, the Poisson's coefficient attains more extreme values because the foldings of the two curves $\gt\alpha$ and $\gt\beta$ become increasingly decorrelated (Fig.~\ref{fig:2}, bottom). \comment{Finite element simulations carried over thin elastic shells confirm the theoretical findings; see~\cite{Note1}.}

The above formulae can be generalized in a straightforward fashion to curved creases $\gt \alpha: u\mapsto (u,f(u),0)$ with arbitrary profile $f$, be it smooth or piecewise smooth, as long as $\abs{f'}<\infty$. In~\cite{Lee2018}, Lee et al. investigated folded states where $f$ described an elastica curve, but neither the folding motion $\t x^\theta$, nor the Poisson's coefficient $\nu$ were provided. Closer to the foregoing investigations is the work of Mundilova~\cite{Mundilova2019} (see also~\cite{Mundilova2018}). Mundilova computed a one-parameter family of isometries that deform a planar domain into a cylindrical domain while preserving the planarity of a given curve, e.g., $\t y^\theta:(u,v)\mapsto \gt\alpha^\theta(u)+v(0,\cos\theta,\sin\theta)$. Then, $\t x^\theta$ is recovered by composing $\t y^\theta$ with a sequence of reflections about two planes, one containing the planar curve and another parallel to it, say, $\{z=0\}$ and $\{z=\sin\theta\}$. In contrast to this geometric approach, the present analytical approach has the advantage of carrying over to other non-developable surfaces of translation, i.e., where the ``facets'' are doubly curved. Fig.~\ref{fig:S} illustrates the reach of the approach over a number of examples; the detail can be found in the Supplemental Material~\footnote{Supplemental material is available here. Code for simulation and figures is available at \url{https://github.com/nassarh/curvedCreaseOrigami}.}

\begin{figure}
    \centering
    \includegraphics[width=\linewidth]{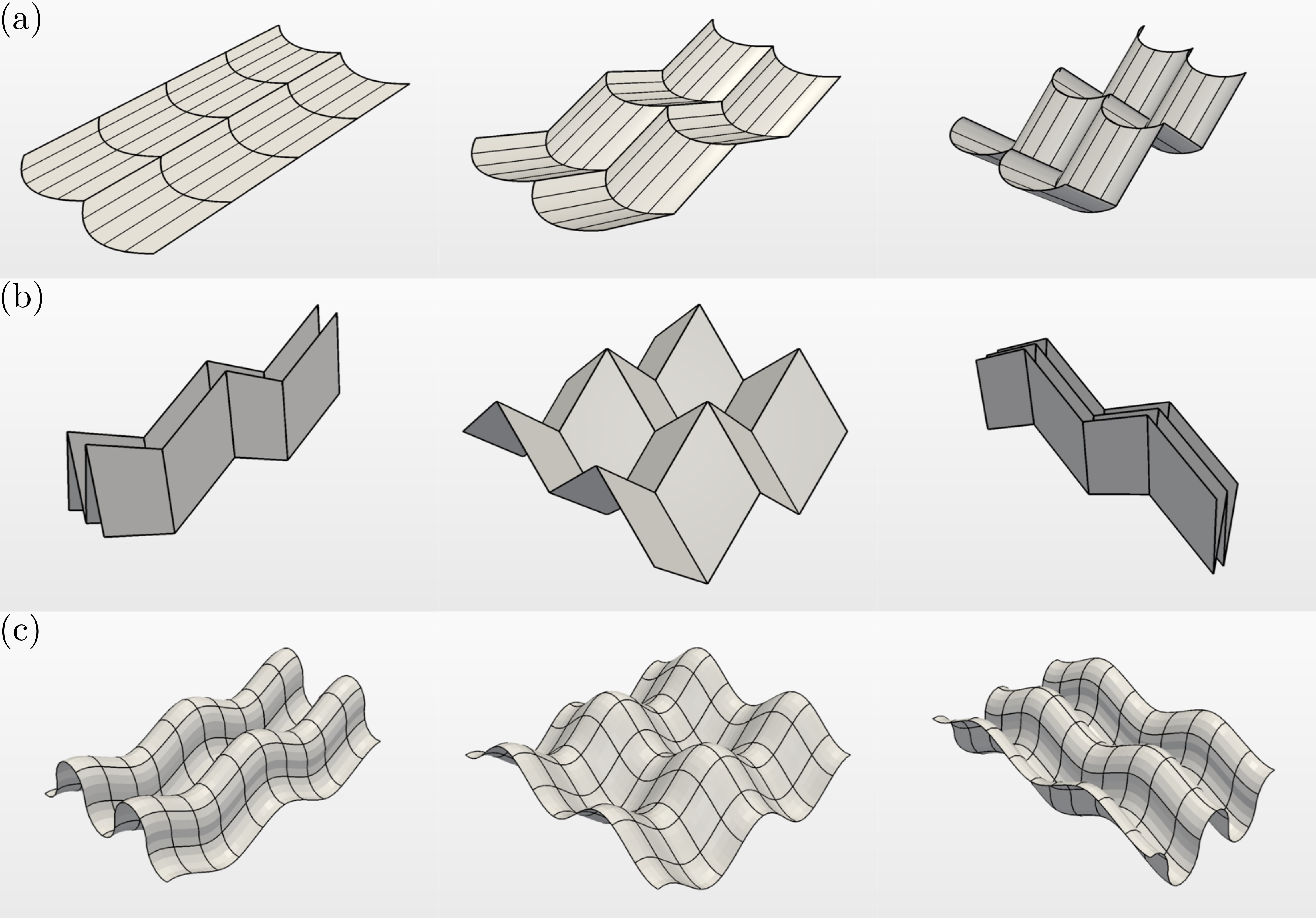}
    \caption{Isometric deformations of surfaces of translation: (a) a surface with both straight and curved creases; (b) a non-developable surface with straight creases known as the ``eggbox'' pattern; (c) a creaseless smooth variant of (b).}
    \label{fig:S}
\end{figure}

\begin{figure}
    \centering
    \includegraphics[width=\linewidth]{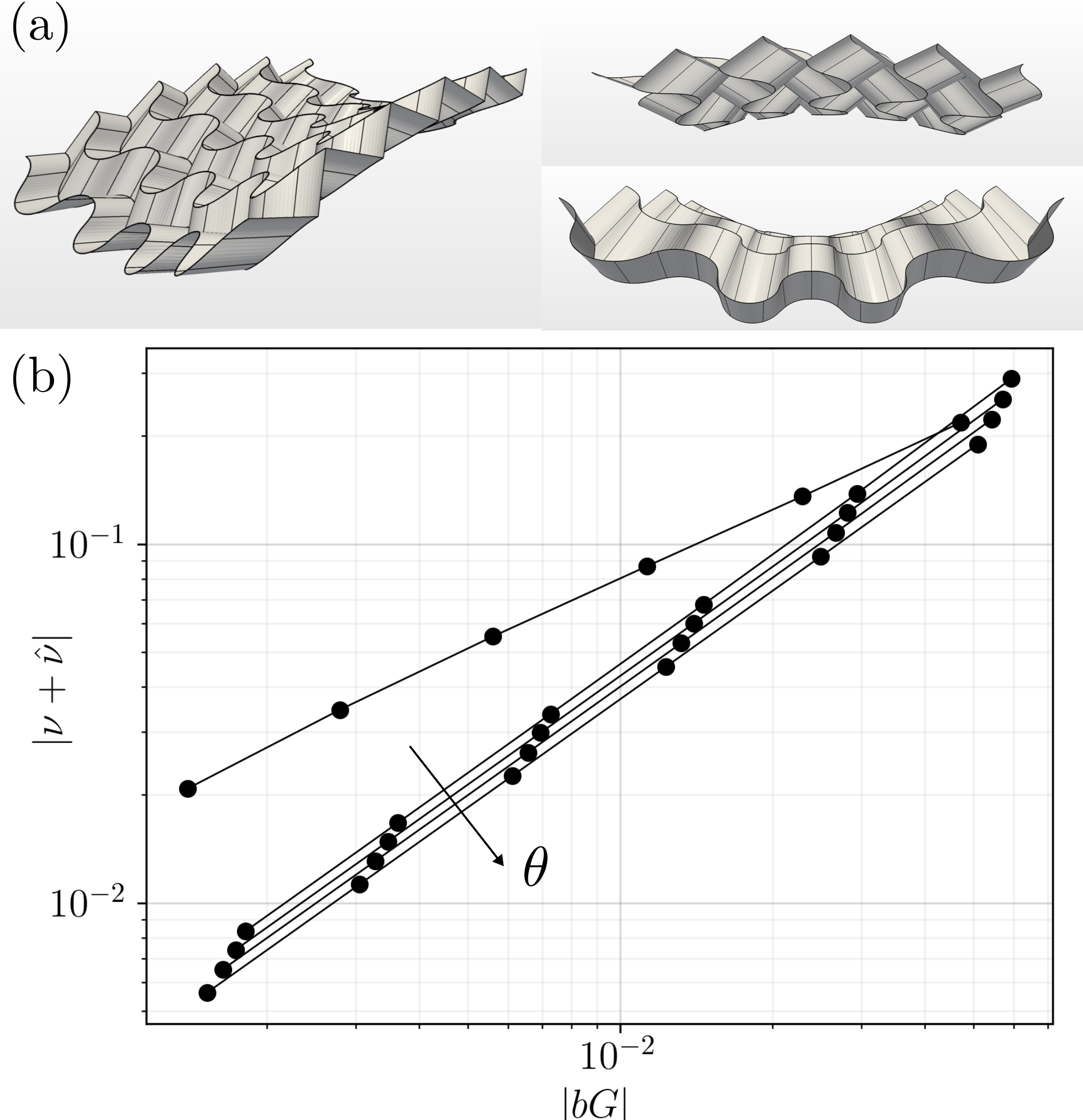}
    \caption{Normal curvatures. (a) An infinitesimally bent state: top (resp. bottom) view puts forward the imposed (resp. induced) curvature in the zigzag (resp. curved crease) direction; inset highlights the saddle shape. (b) Identity of Poisson's coefficients: numerical error v.s. normalized curvature for various folding angles $\theta$ sampled every $\sim5^\circ$ from $\sim22^\circ$ to $\sim45^\circ$; $\theta$ increases with the arrow except for the outlier which corresponds to an almost maximally folded state (for $\theta/\Theta\sim 1$, $\nu$~changes drastically; see Fig.~\ref{fig:2}. This justifies the slower convergence rate).}
    \label{fig:3}
\end{figure}

There is another purely geometric coefficient that characterizes the response of origami tessellations, in bending this time. To introduce it, let $\t x^\theta$ be a folded state and consider an infinitesimal displacement of the form
\[
\t D: (u,v) \mapsto (x(u,v),y(u,v),z(u,v)+Eu^2/2+Gv^2/2).
\]
This displacement is meant to describe an effective bending displacement of $\t x^\theta$ that induces given normal curvatures $\kappa_1\equiv E/I_{11}^\theta$ and and $\kappa_2\equiv G/I_{22}^\theta$. Then, the ratio of normal curvatures, namely $\hat \nu \equiv -\kappa_2/\kappa_1$, mimics an ``out of plane'' Poisson's coefficient and is worthy of investigation: on one hand, its sign indicates whether the tessellation bends into a dome or a saddle; on the other hand, its magnitude quantifies the discrepancy, or lack thereof, between the curvatures in the two directions of periodicity. To find $\hat \nu$, one must find $\t D$ such that $\t x^\theta + \t D$ is infinitesimally isometric to $\t x^\theta$. For the curved-crease variant of the Miura-ori under consideration, we carried numerical simulations where $\theta$ is given, $G$ is imposed at a boundary, say $\{u=0\}$, $\t D$ is computed by propagating (infinitesimal) inextensibility constraints away from the boundary~\cite{Nassar2018b, Note1} and $E$ is evaluated by finite differences. The infinitesimally bent state is illustrated on Fig.~\ref{fig:3}a and the the numerical results are compiled in Fig.~\ref{fig:3}b. These show that
\[\label{eq:id}
\hat \nu = -\nu,
\]
up to a small error proportional to $bG$~\footnote{The results are valid for $G \ll 1/b\sim q$: either bending is finite and the crease pattern is refined; or: the crease pattern is fixed and bending is infinitesimal.}. This identity was first discovered for the original Miura-ori~\cite{Schenk2013, Wei2013} and later noted for other patterns with straight creases (e.g., the ``eggbox'' pattern~\cite{Nassar2017a}, the ``morph'' pattern~\cite{Pratapa2019} and ``zigzag sums''~\cite{Nassar2022}). A recent theoretical development shows that it holds for any surface of translation~\cite{Nassar2023}. Identity~\eqref{eq:id} shows that the in-plane and out-of-plane responses of curved-crease origami tessellations are tied together. Although this can be a hindrance for design, e.g., free form, purposes, it can be a major advantage for control purposes: by imposing one curvature along a boundary, the geometry in the bulk of the tessellation's domain can be specified.

\begin{figure*}
    \centering
    \includegraphics[width=\linewidth]{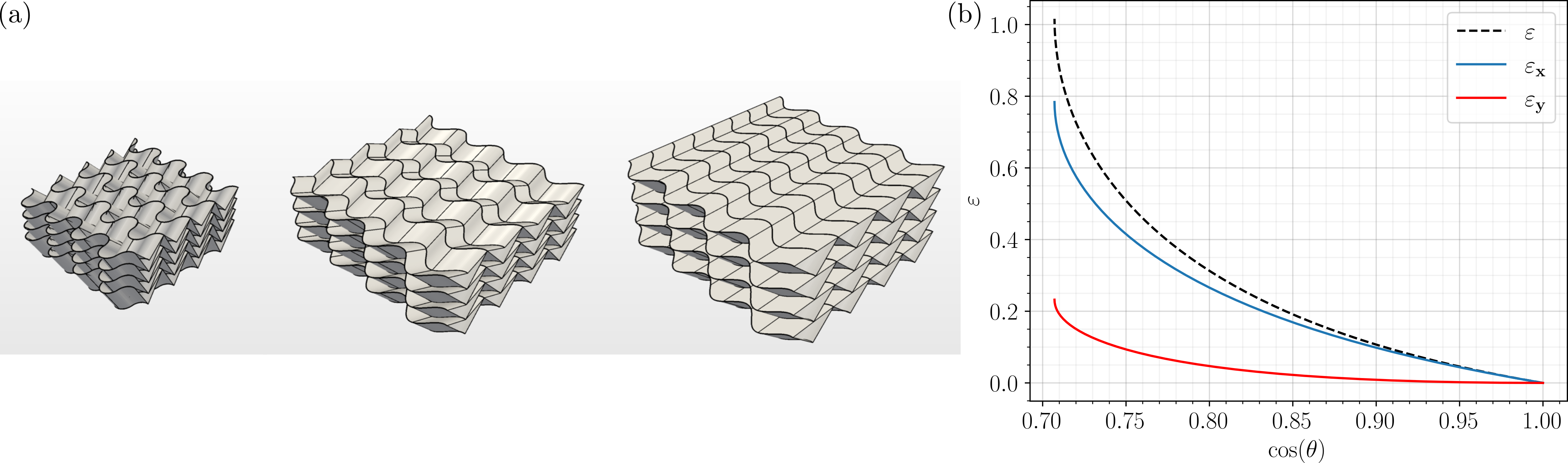}
    \caption{Curved-crease morphing solid. (a) Left to right are maximally folded \comment{($\theta=\Theta=\pi/4$, $\delta=\pi/12$)}, partially folded, and maximally unfolded \comment{($\theta=\delta=0$)} states. \comment{Here, $\theta_o=\pi/4$; other values are the same as for Fig.~\ref{fig:1}}. \comment{(b) Normalized elastic energy ($\str$) v.s. elongation ($\cos\theta$).}}
    \label{fig:4}
\end{figure*}

As discussed above, upon folding, the curved crease $\gt\alpha^\theta$ remains planar. Therefore, it is possible to produce a 3D foldable solid by stacking together copies of $\t x^\theta$ and of its mirror image relative to $\{z=0\}$, in an alternating fashion. Following Schenk and Guest~\cite{Schenk2013}, it is possible to assemble more interesting foldable solids with a self-locking property. Instead of stacking copies of $\t x^\theta$, consider another curved-crease origami $\t y$ given by\comment{
\[
    \begin{split}
    \t y_u(u) &= (1, aq\cos(q u), 0),\\
    \t y_v(v) &=
    \begin{cases}
         (0,\cos\theta_o, \sin\theta_o) \quad &\text{for} \quad 0\leq v < c,\\
        (0,\cos\theta_o, -\sin\theta_o) \quad &\text{for} \quad c\leq v < 2c.
    \end{cases}
    \end{split}
\]}%
Note that $\t y$ is pre-folded through an initial angle $\theta_o$ equal to the initial elevation of the zigzag lines $\{u=cst\}$ of $\t y$ above the horizontal. Thus, it is possible to assemble $\t x$ and $\t y$ by gluing them, say, at even numbered creases $\{v = 2mb\}$ and $\{v = 2mc\}$ if the steps $b$ and $c$ are such that $b = c\cos\theta_o$. As $\t x$ folds into $\t x^\theta$, $\t y$ folds into a surface $\t y^\delta$ defined to be isometric to $\t y$ in the same manner presented above, namely with
\[
    \begin{aligned}
        \t y^\delta_u(u) &= 
        \left(
        \sqrt{1+a^2q^2\cos^2(qu)\left(1-\frac{\cos^2\theta_o}{\cos^2(\theta_o+\delta)}\right)},\right.\\
        &\qquad\qquad\left.aq\frac{\cos(qu)\cos\theta_o}{\cos(\theta_o+\delta)}, 
        0
        \right),\\
        \t y_v(v) &= (0,\cos(\theta_o+\delta),(-1)^m\sin(\theta_o+\delta)),
    \end{aligned}
\]
where the deformation parameter $\delta$ is chosen, if possible, to ensure that contact between $\t x^\theta$ and $\t y^\delta$ is maintained at the same even numbered creases. Comparing the expressions for $\t y^\delta_u$ and $\t x^\theta_u$, it turns out that this is possible if
\[
    \cos(\theta_o+\delta)= \cos\theta \cos\theta_o.  
\]
By alternating copies of $\t x^\theta$ and of $\t y^\delta$, a foldable solid is obtained: the solid is maximally folded for $\theta=\Theta$ then, as $\theta$ decreases, it unfolds up to $\theta=0$. At that point, $\t x^{\theta=0}=\t x$ is flat-unfolded and brings the unfolding of the solid to a halt. This motion is illustrated in Fig.~\ref{fig:4}a. Such solids with predetermined unfolded states provide deployable sandwich panels with curved-crease foldcores for space structures and morphing airfoils applications. Note that the maximally unfolded ``locked'' state can be tuned by adjusting the pre-folding parameter $\theta_o$. Note also that the Poisson's coefficients in $(x,z)$ and $(z,y)$ planes can be computed using the expressions of $\t x^\theta$ and $\t y^\delta$ but this is not pursued here.

\comment{Part of the energy required to deploy the solid is due to crease folding and can be modeled as a lineal density function of dihedral angles at crease lines; this density promises to be highly nonlinear as well as material and history dependent except in a limiting case where the creases have zero stiffness~\cite{Grey2020}. This ideal limit is adopted here and our focus shifts to the remainder of elastic energy due to bending in between the creases. The fact that the deformation is isometric implies that bending energy~$\str\equiv\str(\theta)$ is function of a single measure of curvature, e.g., the mean curvature~$H$ should the constitutive material be isotropic. Thus, up to a coefficient of bending rigidity, $\str = \str_{\t x} + \str_{\t y}$ with
\[
\str_{\t x}(\theta) = \frac12\int_0^{2\pi/q}\int_0^{2b} \left[H(u,v;\t x^\theta)-H^0(u,v)\right]^2\dd u\dd v
\]
being the bending energy of layer $\t x$; $H^0$ is its mean curvature in the natural configuration; and, $H$ is its mean curvature~\cite{Note1}. Energy $\str_{\t y}$ is similarly defined. The profiles of $\str$, $\str_{\t x}$ and $\str_{\t y}$ v.s. elongation as measured by $\cos(\theta)$ are depicted on Fig.~\ref{fig:4}b assuming that the natural configurations of the layers are the ones that occur in a maximally deployed solid. Interestingly, $\str$ reaches the natural state with a non-zero slope; this slope is equal to the buckling load required to trigger the folding of layer $\t x$ out of its flat-unfolded state.}

In conclusion, the presence of curved creases couples bending and folding and requires new methods for the design and analysis of deployable structures, compliant shell mechanisms and morphing metamaterials. The present paper contributes to that effort for a class of curved-crease variants of the Miura ori, as well as other developable and non-developable surfaces of translation. Although the methods proposed here are specific to these cases, they provide insight into the behavior of more general curved-crease origami patterns and a useful benchmark for the development of more universal theories and tools of numerical simulation. \comment{Analysis here was restricted to pre-defined deployment paths; a more general description needs to fully account for the interaction between the inextensible kinematics and the elastic energy landscape.}

\begin{acknowledgments}
Work supported by the NSF under CAREER award No. CMMI-2045881.
\end{acknowledgments}

\end{document}